\documentclass[epj]{svjour}
\usepackage{latexsym}
\usepackage{graphics}
\begin{document}
\title{Saddle-point energies  
and Monte Carlo simulation of the long-range order relaxation in CoPt.}
%\subtitle{Do you have a subtitle?\\ If so, write it here}
\author{M. Allalen\inst{1,}\inst{2}\and H. Bouzar\inst{2} \and T. Mehaddene\inst{3} }                     % Do not remove
\offprints{ M. Allalen}          % Insert a name or remove this line
\institute{Universit\"at Osnabr\"uck, Fachbereich Physik,
D-49069 Osnabr\"uck, Germany \and LPCQ, University M. Mammeri, 15000 Tizi-Ouzou, Algeria   \and Physik Department E13/FRM II, TU M\"unchen, 
 D-85747 Garching, Germany 
 }
\date{Received: date / Revised version: date}
% The correct dates will be entered by Springer
%
\abstract{
We present atomic-scale computer simulations in equiatomic L1$_0$-CoPt where  Molecular Dynamics and Monte Carlo
 techniques have both  been applied to study the vacancy-atom exchange and kinetics relaxation. The atomic potential is determined 
using a Tight-Binding formalism within the Second-Moment Approximation. 
% so as to reproduce correctly 
%the lattice parameters and the formation energies of the ordered and disordered phases of CoPt. 
It is 
used to evaluate the different saddle-point energies  involved in a vacancy-atom exchange between nearest-neighbour sites. The potential and the saddle-point energies have been used to simulate the relaxation of the long-range order in CoPt using a Monte Carlo technique. A vacancy migration energy of $0.73\pm 0.15~eV$ and an order-disorder transition temperature of 935~K have been found. 
\PACS{
      {61.43.Bn}{Structural modeling: computer simulation}   \and
      {64.60.Cn}{Order-disorder transformations} \and
      {66.30.Fq}{Selfdiffusion in metals and alloys}
     } % end of PACS codes
} %end of abstract
\maketitle

\section{Introduction}
\label{sec-1}
 Ordering kinetics in intermetallic compounds have been the topic of many experimental and theoretical works. Systems presenting a phase diagram derived from the Au--Cu canonical phase digram are among the most studied. Owing to their high magnetic anisotropy, some of them in the iron-group metals and platinum group metals  are of outstanding technological importance \cite{1,2,3,4,5}. 
 A good knowledge of the ordering process and its dynamics is thus a necessary step in any extensive research on these systems. Among the energetic parameters that drives diffusion, the saddle-point energies and the migration energies are generally less known and hard to measure in ordered intermetallic compounds. In pure metals and random alloys, the migration energy can be deduced, for example, from stage III of resistivity recovery during annealing after low-temperature irradiation \cite{Schultz} or thorough analysis of residual resistometry along isothermal and isochronal annealing series \cite{Balanzat,Schulze}. Those two methods are, however, very sensitive to the microstructure of the samples and to any impurities or defects. Based on the earlier work of Flynn \cite{flynn}, Schober {\it et al.} \cite{schober} have proposed a model for determining the migration energy from the phonon density of states in pure fcc and bcc metals. This model has been extended latter on to AB$_3$ compounds with L1$_2$ structure \cite{kentz1}. A prerequisite for such a determination is the knowledge of the phonon dispersion at different temperatures and states of order. In parallel to these experimental approaches, computer simulations, based on Molecular Dynamics (MD) and Monte Carlo (MC) techniques,  
 allow one a wide range of kinetics relaxation problems and thermodynamic properties to be studied. 
 Offering a possibility for investigating atomic migration in terms of crystal energetics treated by means of advanced solid-state theories, MD simulations are, however, technically limited to rather small samples and short periods of real time. Consequently, being very powerful when considering the dynamics of single atoms in a solid \cite{massobrio}, the method is less useful for studying net kinetic effects such as, e.g., long-range order (LRO) relaxation. Most of simulation studies of structural kinetics are thus carried out by means of the MC technique (for references see numerous works of Binder, e.g., Ref. \cite{binder}).
Crucial for the success of any simulation, the interatomic potential can be deduced from first-principle MD, providing an accurate
description of the atomic interactions, but requiring enormous computational time and a limited number of particles. To a great extent, this limitation can be overcome by using empirical or semi-empirical potentials which have the advantage of reproducing fast and with 
satisfying accuracy the thermodynamic  and structural properties of materials. Very satisfying results have been obtained, in this way, in transition metals and alloys \cite{Cleri,Rosato}.

The Co-Pt system has interested many researchers from a practical viewpoint as providing catalyst materials \cite{zyade}, but from a basic viewpoint as a magnetic system with a coupling between chemical and magnetic ordering \cite{xiao,cadeville,sanch}. Around the 50/50 stoichiometry,  CoPt orders in the L1$_0$ structure, made of alternating pure cobalt and platinum (100) planes \cite{Massalski}. This very anisotropic chemical order is accompanied by a strong magnetic anisotropy and a tetragonality ($c/a < 1$).  The pronounced anisotropic properties of CoPt are at the core of the present renewed interest in this system \cite{Uba,Maret,Wantanabe}. In thin films, L1$_0$-CoPt alloys display, in addition to a high magnetic anisotropy, a magnetisation that is perpendicular to the film surface , thus making such films good candidates for magneto-optical-storage devices \cite{Shiomi}. This system has been studied extensively to understand the origin of the asymmetry in its phase diagram. Various explanations have been proposed: for example, an effect of many-body interactions \cite{sanch}, an influence of magnetism, and variations of atomic interaction with composition \cite{17,18}. Calculation of this phase diagram may appear quite simple {\it a priori}, as it is based on the fcc lattice at almost all temperatures and compositions, but it still remains a challenge today.  The kinetics of atomic ordering in the ordered phases in this system has also been studied. The activation energy of changes in chemical LRO has been determined by resistivity measurements to be $3.18~eV$ and $2.52~eV$ for CoPt$_3$ \cite{21} and CoPt \cite{22} respectively.     

It is well known, that the vacancy-driven mechanism, namely, the  vacancy-atom jump between nearest-neighbours  sites, 
 is the dominant microscopic process of diffusion in dense phases \cite{Petry}. During  such a jump, the energy barrier formed by the nearest-neighbours that the jumping atom has to overcome for completing the jump plays an important role. In this paper, we present MD simulations based on the Tight-Binding  Second Moment Approximation (TB-SMA) formalism \cite{Tomanek}, which is well adapted to transition metals and alloys \cite{duca}, to calculate the saddle-point energies for the different kinds of atomic jumps. The calculated saddle-point energies  will be used, in conjunction to the potentials deduced from the TB-SMA, to simulate the LRO relaxation in CoPt and deduce the migration energy using a MC technique.

\section{Saddle-point energies calculation}
 In this section, we present the many-body potential and its parametrisation procedure. Once  well defined, the potential will be used to simulate the atom-vacancy jump mechanism and deduce the energy barriers for the different kinds of atomic jumps to a nearest-neighbour vacancy in the L1$_{0}$-CoPt structure in the case of a rigid and a relaxed lattice.

The potential we have used is based on the approach proposed by
 Rosato,
Guillope, and  Legrand \cite{Rosato,Guillope,Treglia} where the energy $E_{i}$ of an atom at site $i$, derived
in the  TB-SMA formalism \cite{Tomanek}, is written as
the sum of two terms; an  attractive band energy ($E_{i}^{b}$)
 and a repulsive pair interaction term ($E_{i}^{r}$).
The band term is obtained by integrating the local density of states up to the Fermi level
\cite{Friedel}, this gives rise to the many-body character of the potential necessary 
to account for surface relaxations and reconstructions whereas the repulsive term  is described by a sum of  Born-Mayer ion-ion repulsions.
 When replacing the realistic density of states by a schematic rectangular one having 
the same second-moment  \cite{Rosato}, one obtains:

\begin{equation}
E_{i}^{b}=- \sqrt{\sum_{j,r_{ij}<r_{c}} \xi^2_{IJ} exp[-2q_{IJ}(\frac{r_{ij}}{r^{IJ}_{0}}-1)]}
\end{equation}
and
\begin{equation}
E_{i}^{r}= \sum_{j,r_{ij}<r_{c}}A_{IJ} exp[-p_{IJ}(\frac{r_{ij}}{r^{IJ}_{0}}-1)]
\end{equation}
I and J indicate
the chemical species Co or Pt.
 $\xi_{IJ}$ is an effective hopping integral and q$_{IJ}$ describes its dependence on the relative interaction distance. p$_{IJ}$ is related to the bulk modulus of the alloy under consideration. In our 
case $r_{0}^{II}$ is the first-neighbour distance
in the metal I. $r_{0}^{IJ}$  as a free parameter and it is
 different from the first-neighbour distance r$_{0}^{IJ}$=( r$_{0}^{II}$+
 r$_{0}^{JJ}$)/2. r$_{ij}$ is the distance between atoms at sites
$i$ and $j$. $E_{i}^{b}$ and $E_{i}^{r}$  
are cancelled beyond a cut-off radius r$_{c}$ which has been chosen as the second-neighbour distance. From r$_{c}$ up to the (Co-Co) third-neighbour  distance, the  potential is linked up 
to zero  with a fifth-order polynomial in order to avoid
discontinuities, both in the energy and in the forces. The TB-SMA potential  applied to intermetallic compounds gives very reliable results with interactions up to the second nearest-neighbour as long as intermediate and low-temperature ranges are considered. Long-range interactions are required in order to reproduce high-temperature properties which are usually beyond the capability of short-range potentials. Considering the temperature range (600-980~K) and the goal of our simulations, the approximation of interactions up to the second nearest-neighbours is rather justified. Furthermore, x-ray and neutron diffuse scattering measurements have been used in conjunction with inverse cluster variation method to calculate the effective pair interactions in L1$_{2}$-Co$_{3}$Pt and in L1$_{2}$-CoPt$_{3}$ \cite{17,18}. In both systems the first and second interactions are the predominant ones. The same calculations are now in progress in equiatomic CoPt \cite{hamidi}.

 The potential parameters $\xi_{IJ}$, A$_{IJ}$, p$_{IJ}$ and  q$_{IJ}$ are {\it a priori} unknown and will be determined by fitting
the potential to the universal equation of state driving the variation of the potential
with distance \cite{Cleri}. This procedure requires usually the knowledge of the cohesive energy,
lattice parameters and bulk modulus of the system. We have taken these parameters from the literature \cite{Goyhenex}. We aimed during the fit 
 to reproduce correctly the formation energies and the lattice parameters of the ordered L1$_{0}$ and the disordered A$_{1}$ phases of CoPt
. 

 The final  parameters of the potential are summarised  in Table I. In Table, II we report  together the calculated lattice parameters and the  formation  energies and compare them to the experimental data. A good agreement is achieved for both ordered L1$_{0}$ and  disordered A1 phases.
 We have attempted to parameterise the TB-SMA potential for the L1$_2$-CoPt$_3$ phase but a discrepancy of more than 40\% appeared between the experimental and the  calculated formation energies. For this reason, further calculations, aiming at the determination of the saddle-point energies, have been restricted to the  L1$_0$ phase.

In order to calculate the saddle-point energies, we have used a simulation box containing 10$^{3}$ L1$_0$ fcc-based cells. In accord to the L1$_{0}$ structure, there are as many Co as Pt atoms arranged on (100) alternating crystallographic planes of pure Co and Pt using
periodic boundary conditions in the three space directions. The total energy of the system, containing a single vacancy, was monitored during the
four kinds of atomic jumps to a nearest-neighbour vacancy in the case of a rigid and a relaxed lattice:
 Co to Co-site vacancy (Co$\rightarrow$Co(V)), Co to
Pt-site vacancy (Co$\rightarrow$Pt(V)),  Pt to Pt-site vacancy (Pt$\rightarrow$Pt(V)) and Pt to Co-site vacancy (Pt$\rightarrow$Co(V)). In Fig.~1 we represent the two possible jumps of a Pt atom toward a vacancy 
situated either in the same plane (Pt$\rightarrow$Pt(V)) or  in a neighbouring plane 
( Pt$\rightarrow$Co(V)).   
In the case of a relaxed lattice, we have used  a Tight Binding-Quenched Molecular Dynamics-Second Moment Approximation (TB-QMD-SMA), which 
allows us to determine  the equilibrium structure of a system
 with a finite number of particles at $T=0~K$, by integrating the equation of motion \cite{Bennett1}.
The quenching procedure, in which the velocity $v_{i}$ of an atom $i$  is cancelled when the product $F_{i}(t)$$v_{i}(t)$
is negative, leads to the minimisation of the potential energy at $0~K$ \cite{Bennett2}. $F_{i}$ being the force acting on the atom $i$, calculated in the extended tight-binding formalism from the
total energy \cite{Friedel}. The positions were integrated by means of the Verlet algorithm \cite{Verlet}

 For all atomic jumps, the total energy follows a smooth cosine curve, with a maximum close to  half-way between the initial and destination 
lattice sites. In the case of the cross jumps, Co$\rightarrow$Pt(V) and Pt$\rightarrow$Co(V), a shift in the total energy, associated to the formation of an anti-site defect, was observed at the final position.  
The saddle-point energies have been deduced from the evolution of the total energy for the different kinds of atomic jumps. The results obtained with and without relaxation are reported in Table III. The saddle-point energies are denoted $\epsilon^R$ and $\epsilon$ for the relaxed and the non-relaxed lattice respectively. The amplitude of the saddle-point energies may be explained by the size effect and the atoms forming the barrier of first-nearest neighbours that the jumping atom has to  overcome for completing the jump. For instance, the largest value is obtained in the case of Pt$\rightarrow$Pt(V) jump, when the bigger Pt atom is involved in the jump. In the case of the cross jumps,  Co$\rightarrow$Pt(V) and Pt$\rightarrow$Co(V), the atomic structure of the barrier to overcome being the same for the two jumps, the corresponding saddle-point energies are very close. Furthermore, the lattice being softer when relaxed, the saddle-point energies, are as expected, smaller when the relaxation is taken into account.

\section{Monte Carlo simulations}
\subsection{Simulation method }
MC simulations have been established as a useful tool for studying order-order and 
order-disorder relaxation kinetics in intermetallics \cite{Oramus,Oramus1,A.Kerrache}.
In contrast to earlier studies, based on effective pair interaction energies expressed within a simple Ising-Hamiltonian, we go in the present work beyond the pair approximation by implementing the many-body potential deduced from the TB-SMA and taking into account the saddle-point energies calculated in the previous section to simulate the disordering process of a perfectly ordered L1$_{0}$-CoPt structure. The purpose of
the simulation is to prove whether the interaction model is appropriate and to get an estimation of the vacancy migration energy which is an important energetic parameter that drives diffusion and ordering process in intermetallics. 

In view of recent neutron diffuse scattering measurement in equiatomic CoPt \cite{hamidi} which show a symmetric distribution of the diffuse intensity around the 100 and equivalent points in the reciprocal lattice, signature of a highly stable L1$_0$ phase with very small static displacements, we have chosen to carry on the MC simulations using a rigid lattice.  
We have used a model  based on the vacancy-atom jump mechanism between nearest-neighbour  sites,
which is the realistic microscopic process in dense phases
\cite{Petry}.
The simulation box contains $32^{3}$ L1$_0$ fcc-based cells with linear periodic boundary 
conditions. The simulation starts with a perfect L1$_0$ ordered crystal in which 
one of the two sublattices (sublattice $\alpha$) is occupied by Co atoms and the 
other (sublattice $\beta$) by Pt atoms.  To not affect the static properties and avoid interaction effects, a single vacancy is introduced at random in 
the crystal. 
 The elementary MC step is the following: one of
the vacancy neighbours (Co or Pt atom) is randomly chosen, the energy balance $\Delta E$ of the atom-vacancy exchange 
 before and after the
jump is evaluated.  The jump  is performed if the Glauber probability
$P(\Delta E)=P^{G}(\Delta E)exp(-\epsilon/k_{B}T)$ is larger than a random number between
$0$ and $1$. $P^{G}(\Delta E)=[1+exp(\Delta E/k_{B}T)]^{-1}$ is the Glauber probability \cite{Glauber} and $\epsilon$ is the saddle-point energy for the corresponding jump.  This corresponds to averaging the result over a large number of reversal jump
attempts, the sum of the probabilities of the jump and its  reversal being equal to $1$.
 
The configuration of the system was analysed, at regular time intervals, 
by calculating the  LRO parameter $\eta=2(2N_{\rm Co}^{\alpha}-N_{\rm Co})/(N_{\rm sites}-1)$, 
where $N_{\rm Co}^{\alpha}$ is the number of Co atoms on the $\alpha$ sublattice and
$N_{\rm Co}$ the total number of Co atoms. The time scale being the number of jump attempts. For each temperature, the evolution of  $\eta$  
is followed until the system reaches equilibrium. We have chosen to 
stop when the simulation time was at least longer than 5 times the 
 relaxation time of the system.

\subsection{Results and discussion}
Isothermal relaxations of $\eta$ have been recorded in the temperature range from $600$ to $980~K$. An example of kinetics relaxation  
is shown in Fig.~2. According to the path probability method \cite{ppm}, corroborated by many experimental  results  
\cite{30,31,32,33,kulovits}, the isothermal relaxation of the LRO parameter in intermetallics is well fitted with two exponentials, yielding a long and a short relaxation times, corresponding to the slow and the fast processes, respectively.  This result is well established in L1$_{2}$ phase for which a detailed study  has shown that the fast process 
is related to the formation of the nearest-neighbour antisite pairs, whereas the 
slow one is related to the uncoupling of these antisite pairs \cite{Oramus}.  
Due to the difference in structure, the process must be different in the L1$_0$ 
phase and the interpretation is still under investigation. However, recent MC simulations of the LRO relaxation in L1$_0$-FePd showed that the fast process is highly predominant (up to 90\%) below the order-disorder transition \cite{scripta}. In view of these results and considering the temperature range of our simulations, we have chosen to fit the kinetics using a single exponential yielding a single relaxation time $\tau$ which fulfils an Arrhenius law with a positive migration energy $E_{M}=0.73\pm 0.15~eV$
 (Fig.~3). The relaxation times are typically in the order of 10$^{12}$ MC steps, significantly higher than the relaxation times obtained without saddle-point energies. The  data points on the left side of Fig.~3 , plotted in open circles, show a clear departure from the straight line, they are synonym of a critical slowing down close to the order-disorder transition and have been excluded from the linear regression. 

The vacancy formation energy E$_{F}$ has been measured in pure Co and pure Pt, it has been found equal to $1.38~eV$ and $1.2~eV$ for Co and Pt respectively \cite{E_F}. Assuming that the activation energy E$_A$ is the sum of E$_{M}$ and E$_{F}$, the calculated value of E$_A$ obtained considering the simulated value of E$_{M}$ and an average value of E$_{F}$  compares well to the activation energy of $2.52~eV$, measured in equiatomic CoPt by means of resistivity measurements \cite{22}.  Normal modes of vibration in equiatomic CoPt have been recently measured  by inelastic neutron scattering \cite{messad2}. The phonon density of states has been used to calculate the migration energy using Schober's model. A value of $0.85~eV$ was found in the fcc-disordered state at $1120~K$. Despite of the difference in the state of order between the simulated CoPt system and the measured one, the qualitative agreement is quite satisfactory. In our knowledge no further measurement of the migration energy in equiatomic CoPt are available. Nevertheless, qualitative comparison to CoPt$_{3}$, in which the migration energy was calculated using both Schober's and Flynn's models, can be made. 

Schober's model initially developed for bcc and fcc pure metals \cite{schober} has been extended to the A$_{3}$B compounds with L1$_2$ structure \cite{kentz1} and applied to CoPt$_{3}$ \cite{Mehaddene}.  Averaging the  values of the migration energy over the different kinds of atomic jumps, we get a migration energy of $1.3~eV$ at $300~K$. Flynn's model \cite{flynn} was applied to CoPt$_{3}$  using the elastic constants deduced from the slope of the phonon dispersion curves at the center of the Brillouin zone \cite{Mehaddene}. A migration energy of $0.95~eV$ was found in the fcc-disordered state at $1060~K$. The difference in the amplitude between the calculated  value of the migration energy in CoPt and the values deduced from Schober's and Flynn's models in CoPt$_{3}$ might be explained by the atomic mass and size effects. As expected, an increase in the migration energy with the atomic density is observed between the L1$_0$ and L1$_2$ phases of Co--Pt system. The atomic density is nearly 25\% lower in CoPt than in CoPt$_3$.

The variation of the equilibrium LRO parameter $\eta_{eq}$  as a function of
temperature is shown in Fig.~4. The  order-disorder transition takes place between $930$ and $935~K$.
The transition region has been crossed with $5~K$ temperature steps. 
It should be noted that  the value of this temperature is close to the value 
obtained in the L1$_0$ compounds using phenomenological
pair interaction energies  \cite{Kerrache} but it is 15\%  lower than  
the experimental value of the order-disorder transition temperature $T_{c}$
in equiatomic CoPt ($1110~K$) \cite{Dah85}. Nevetheless, the general agreement between the simulated and the experimental $T_{c}$ is quite satisfactory. In fact, the MC model, which assumes temperature independent potentials and does not take into account anti-phase domains,  only gives an
 estimate of $T_{c}$. Indeed, recent x-ray diffraction and transmission electron microscopy measurements in CoPt showed that the ordering transformation involves formation of antiphase boundaries and twin bands \cite{xiao}.   
\section{ Conclusion} 
\label{sec-5}

An  approach for determining  the parameters  of a many-body potential in equiatomic CoPt using the TB-SMA
 has been presented. The potential  was used to determine the saddle-point energies for the different kinds of nearest-neighbour atom-vacancy jumps in the case of a rigid and a relaxed lattice. The calculated energy barriers have been used together with the many-body potential to simulate the LRO relaxation in L1$_0$-CoPt structure using a MC technique. A vacancy migration energy of $0.73\pm 0.15~eV$
 has been deduced from the Arrhenius plot of the relaxation times.  A lack in experimental data of the migration energy in the ordered state of the equiatomic CoPt made possible only qualitative comparisons to either the disordered state of CoPt data, when available, or to CoPt$_3$ data. Finally, the order-disorder transition temperature  has been determined  from the isothermal relaxation of the LRO parameter. The simulated value is in a satisfactory agreement with experiment.

\vspace*{2cm}

{\small{
This work was supported partly by the Algerian Project ANDRU/PNR3 and by the collaborative
program 99 MDU 449 between the University Mouloud Mammeri of Tizi-Ouzou, Algeria and
the University Louis Pasteur of Strasbourg, France. The authors would like to thank Pr. Treglia, Dr.  Goyhenex and Dr. Pierron-Bohnes who provided us with the initial MD code .}}

\newpage
\vspace*{2cm}
\begin{figure}[h!]
\resizebox{0.9\columnwidth}{!}{%
  \includegraphics{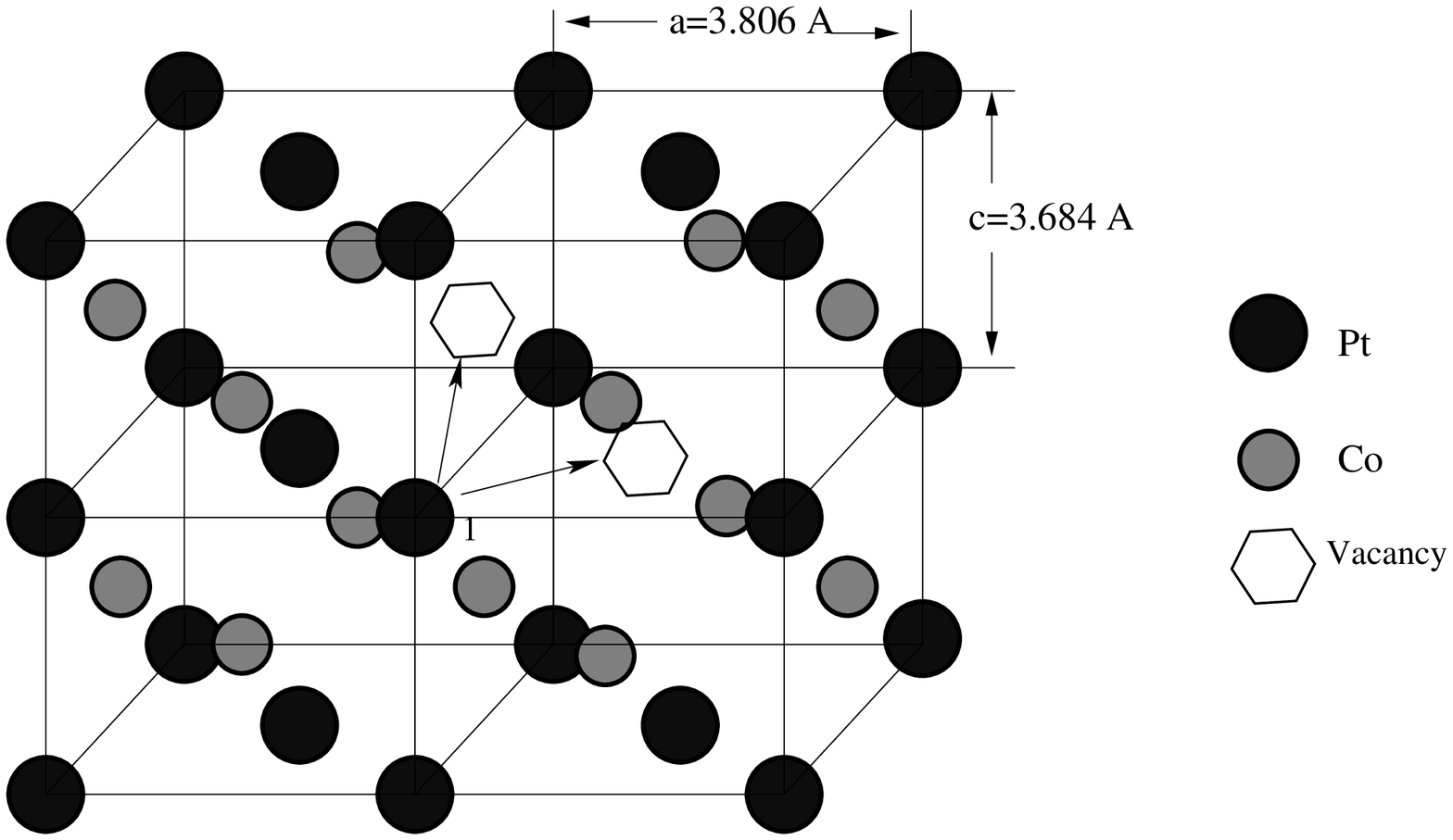}}
\caption{The two possible atomic jumps, in-plane and out-of-plane, 
 of the Pt atom.}
\label{fig2}      
\end{figure}

\newpage
\vspace*{2cm}
\begin{figure}[h!]
\resizebox{0.7\columnwidth}{!}{%
  \includegraphics{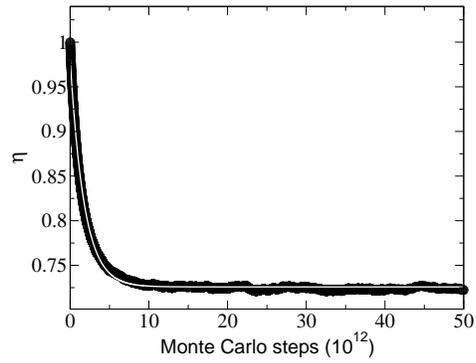}}
\caption{Isothermal relaxation of $\eta$ obtained for $T=880~K$ (black circles) and its simulation using a single exponential (white line).}
\label{fig3}      
\end{figure}

\newpage
\vspace*{2cm}
\begin{figure}[h!]
\resizebox{0.7\columnwidth}{!}{%
  \includegraphics{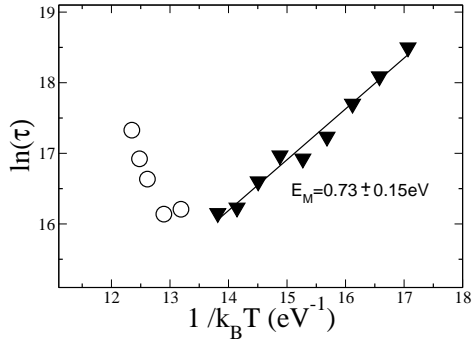}}
\caption{Arrhenius plot of the relaxation times. The linear regression gives
the migration energy $E_{M}= 0.73\pm0.15~eV$. The open circles show the data points which have been excluded from the linear regression.}
\label{fig4}      
\end{figure}

\newpage
\vspace*{2cm}
\begin{figure}[h!]
\resizebox{0.7\columnwidth}{!}{%
  \includegraphics{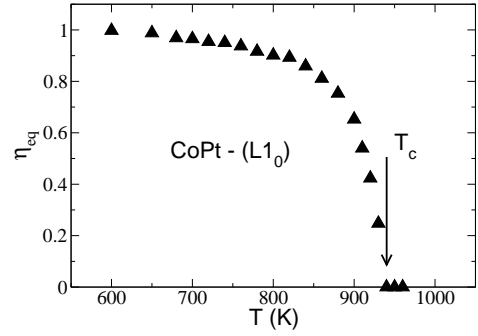}}
\caption{Temperature variation of the equilibrium LRO parameter $\eta_{eq}$.}
\label{fig3}      
\end{figure}

\newpage
\begin{table}
\begin{center}
\caption{ Potential parameters of inter-atomic interactions deduced from the TB-SMA. The interaction parameters between Co-Co and Pt-Pt are taken from the literature \cite{Goyhenex}}
\end{center}
\begin{center}
\begin{tabular}{c|c|c|c|c}
\hline
\hline
&&&&\\
Interaction &A (eV) &$\xi (eV) $ &p &q \\
& & & & \\
\hline
& & & & \\
Pt-Pt&0.242 &2.506 &11.14&3.68 \\
& & & & \\
Co-Co& 0.189 & 1.907 & 8.8&2.96\\
& & & & \\
Pt-Co & 0.175&2.115 & 9.412 & 2.812 \\
& & & & \\
\hline
\hline
\end{tabular}
\end{center}

\end{table}
\begin{table}
\begin{center}
\caption{Reduced lattice parameters, calculated ($E_{cal}$) and experimental ($E_{exp}$) formation energies of the ordered L1$_0$ and disordered A1 phases. In L1$_0$ phase, the results obtained by varying a, c, and both a and c simultaneously are shown. We have used $a_{0}=3.806$~\AA, $c_{0}= 3.684$~$\AA$ for the L1$_0$ phase and $a_{0}=3.751$~$\AA$ for the A1 phase }
\end{center}
\begin{center}
\begin{tabular}{c|cc|cc}
\hline
\hline       
               & $E_{cal}$~(eV)&  $E_{exp}$~(eV) & $a/a_0$& $c/c_0$ \\
\hline
                       &   &&     &   \\
L1$_0$ (varying a)           & -0.133&  -0.14 &0.99&  \\
L1$_0$ (varying c)           & -0.142 & -0.14 && 0.97  \\
L1$_0$ (varying a and c)     & -0.138  &-0.14 &0.995 &0.995        \\
&& &        &   \\
A$_1$  (varying a)        & -0.086&   -0.10    & 1
&                    \\
   \hline
   \hline
\end{tabular}
\end{center}

\end{table}

\begin{table}
\begin{center}
\caption{Calculated saddle-point energies in eV for the different kinds A$\rightarrow$B(V) jumps of an atom A on a vacancy site B in the case of a rigid ($\epsilon$) and a relaxed lattice ($\epsilon^R)$. 
}
\end{center}
\begin{center}
\begin{tabular}{c|cccc}
\hline
\hline       
               & Co$\rightarrow$Co(V)& Co$\rightarrow$Pt(V)& Pt$\rightarrow$Pt(V)& Pt$\rightarrow$Co(V) \\
\hline
$\epsilon$           & 0.49 & 0.41 &0.54& 0.39  \\
 $\epsilon^R$    & 0.39  &0.34 &0.42 &0.31        \\
   \hline
   \hline
\end{tabular}
\end{center}

\end{table}

\end{document}